\documentclass[fleqn,usenatbib]{mnras}

\usepackage{newtxtext,newtxmath}
\usepackage[T1]{fontenc}

\DeclareRobustCommand{\VAN}[3]{#2}
\let\VANthebibliography\thebibliography
\def\thebibliography{\DeclareRobustCommand{\VAN}[3]{##3}\VANthebibliography}

\usepackage{graphicx}	% Including figure files
\usepackage{amsmath}	% Advanced maths commands

\usepackage{xspace}

\let\times\cdot

\newcommand{\spock}{\textsc{spock}\xspace}
\newcommand{\pluto}{\textsc{pluto}\xspace}
\newcommand{\msp}{\textsc{S1}\xspace} % spock model (S1)
\newcommand{\mpg}{\textsc{p-g}\xspace} % pluto model with initial depletion
\newcommand{\mpng}{\textsc{p-ng}\xspace} % pluto model without initial depletion
\newcommand{\msun}{M\ensuremath{_\odot}}

\title[Can photoevaporation open gaps?]{Can photoevaporation open gaps in protoplanetary discs?}

\author[M. L. Weber, B. Ercolano, G. Picogna]{
Michael L. Weber,$^{1,2}$\thanks{E-mail: mweberw@pm.me}
Barbara Ercolano,$^{1,2,3}$, Giovanni Picogna$^{1,2}$
\\
$^{1}$Universitäts-Sternwarte, Ludwig-Maximilians-Universität München,
        Scheinerstr. 1, D-81679 München, Germany\\
$^{2}$Excellence Cluster Origins, Boltzmannstrasse 2, D-85748 Garching bei München, Germany\\
$^{3}$Max-Planck-Institut für Extraterrestrische Physik, Giessenbachstraße 1, 85748 Garching, Germany
}

\date{Accepted XXX. Received YYY; in original form ZZZ}

\pubyear{\the\year{}}


\begin{document}
\label{firstpage}
\pagerange{\pageref{firstpage}--\pageref{lastpage}}
\maketitle

\graphicspath{{figures/}}

\begin{abstract}
We investigate whether photoevaporation alone can open and sustain gaps in protoplanetary discs by coupling the evolving disc structure with the photoevaporative flow in two-dimensional radiation–hydrodynamical simulations. Our results show that once a density depression forms, the local mass-loss rate decreases sharply, suppressing further gap deepening. Viscous inflow and radial mass transport along the disc surface act to partially refill the depleted region, preventing complete clearing. The resulting configuration is a persistent, partially depleted zone whose evolution is largely insensitive to the initial disc morphology. This behaviour challenges the standard paradigm that photoevaporation efficiently carves clean inner cavities and directly produces transition discs. However, the pressure maximum at the outer edge of the depression may still trap dust grains, giving rise to transition-disc–like observational signatures. We also present a first-order prescription to approximate this behaviour in one-dimensional disc evolution models, suitable for use in planet formation and population synthesis studies. Although the prescription improves upon static mass-loss treatments, it remains approximate, underscoring the need for further multidimensional simulations and parameter-space exploration to derive robust recipes for global disc and planet population models.
\end{abstract}


%%%%%%%%%%%%%%%%%%%%%%%%%%%%%%%%%%%%%%%%%%%%%%%%%%

%%%%%%%%%%%%%%%%% BODY OF PAPER %%%%%%%%%%%%%%%%%%

\section{Introduction}
Protoplanetary discs are the birthplaces of planets, as demonstrated by direct detections of forming planets around PDS 70 \citep{Keppler_2018, Muller_2018}. Planet formation and migration proceed within these discs until their dispersal, which occurs rapidly \citep[e.g.][]{Skrutskie_1990} after $\sim2 - 8$\,Myr \citep[e.g.][]{Pfalzner_2022}. Constraining the mechanisms responsible for disc dispersal is therefore essential to understanding the timing and conditions of planet formation.

Photoevaporation -- the removal of gas through thermally driven winds launched by high-energy stellar radiation -- has long been recognised as a key dispersal mechanism \citep[see reviews by ][]{Ercolano_2017b, Lesur_2021, Pascucci_2023}. The process is most efficient at a characteristic radius, typically a few au, where the gas thermal energy exceeds the stellar gravitational potential. When the local mass-loss rate due to photoevaporation exceeds the inward viscous mass flux, a gap can form \citep[e.g.][]{Clarke_2001}. The inner disc then drains onto the star on a short viscous timescale, while the outer disc is eroded from the inside out.
This inside-out dispersal is another important constraint demonstrated by observations\citep[e.g.][]{Strom_1989, Kenyon_1995, Andrews_2005}, including photometric surveys \citep{Koepferl_2013, Ercolano_2011}.

Radiation-hydrodynamical models have confirmed that photoevaporation can produce such gaps and drive disc clearing \citep[e.g.][]{Owen_2010, Owen_2012, Picogna2019, Picogna_2021, Ercolano_2021, Sellek_2024a, Nakatani_2024}. Observationally, this scenario provides a natural explanation for transitional discs -- systems showing inner cavities in infrared and millimetre continuum emission. While many cavities are attributed to planet-disc interactions, a population of low-accretion, cavity-bearing discs is consistent with photoevaporative origin \citep{Ercolano_2023, Garate_2021}.

The theoretical basis for gap opening by photoevaporation remains, however, uncertain. Most models rely on two-dimensional (r–$\theta$) simulations from which one-dimensional radial profiles of the mass-loss rate, $\dot{\Sigma}$(r), are extracted and used as fixed sink terms in viscous evolution calculations \citep[e.g.][]{ercolano2015b, Ercolano_2017, Carrera_2017}. These prescriptions implicitly assume that $\dot{\Sigma}$(r) remains constant during the gap-opening phase, neglecting feedback between the evolving disc structure and the photoevaporative flow. This neglects the fact that once a gap begins to form, local mass-loss rates within the gap region might decrease sharply or vanish, and viscous inflow might refill the transition region, preventing further deepening of the gap. 

If photoevaporation is self-limiting in this way, it may be less efficient than previously assumed at clearing gas at the radii where giant planets form. This has far-reaching implications for models of core accretion, gas envelope growth, and migration, and ultimately for the observed architectures of planetary systems \citep[e.g.][]{Jennings_2018, Monsch_2021}.

In this Letter, we use hydrodynamical simulations to investigate the ability of photoevaporation to open and sustain gaps in protoplanetary discs when the flow dynamically responds to changes in disc structure. We test the validity of the commonly adopted static mass-loss prescription and assess whether photoevaporation alone can account for gap formation and subsequent clearing under realistic conditions.

\section{Methods}
To reduce the computational cost of the simulations, a two-phased approach was employed.
\subsection{Phase I: One-dimensional disc evolution}
In the first phase, the one-dimensional viscous evolution code \spock \citep{ercolano2015b} was used to simulate the evolution of a viscously accreting disc undergoing X-ray photoevaporation until the onset of gap formation. 
\spock evolves the one-dimensional gas surface density, $\Sigma$, according to
\begin{equation}
\frac{\partial \Sigma}{\partial t} = \frac{1}{R} \frac{\partial}{\partial R} \bigl[3 R^{1/2} \frac{\partial}{\partial R}(\nu \Sigma R^{1/2}) \bigr] - \dot{\Sigma}_W(R,t),
\end{equation}
where $R$ is the cylindrical radius, $t$ the time, $\nu$ the kinematic viscosity of the disc, and $\dot{\Sigma}_W(R,t)$ the photoevaporation profile.
The first term on the right-hand side describes viscous evolution \citep{lynden-bell1974}, while the second term is a sink term representing the mass-loss by photoevaporation, parametrised using the mass-loss profile from \citet{Picogna2019}.
The adopted initial conditions were: stellar mass $M_* = 0.7$\,\msun, disc mass $M_{disc} = 4\times10^{-3}$\,\msun, and aspect ratio $\frac{H}{R} = 0.05$ at $R_1 = 50$\,au. 
A locally isothermal disc with temperature $T \propto R^{-1/2}$ was assumed, resulting in a slightly flared structure with $H \propto R^{5/4}$.
Viscosity was prescribed using the $\alpha$-disc model \citep{shakura1973a}, such that $\nu = \alpha c_s H$, with $c_s$ the sound speed and $H$ the scale height of the disc.
An $\alpha$ value of $5\times10^{-4}$ was adopted.
The computational domain was discretized to a grid of 1024 radial cells, equispaced in $R^{1/2}$, extending from 0.1 to 400\,au. 

\begin{figure}
	\centering
	\includegraphics[width=.48\textwidth]{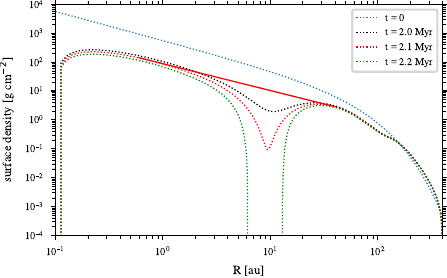}
	\caption[Initial 1D surface density]{
		Surface density evolution in the 1D model \msp. The red dotted line shows the initial conditions for the model with an initial depletion (\mpg). For the model without an initial depletion (\mpng), the surface density was modified using the solid red line, as described in \autoref{sec:met:pluto}.
	}
	\label{fig:sigma0}
\end{figure}
The one-dimensional model was evolved for 2.1\,Myr, at which point a partial depletion in the surface density developed due to the photoevaporative wind, marking the onset of gap formation.
\autoref{fig:sigma0} shows the surface density at that time as well as 100\,kyr before and after.
Throughout this work, the \spock model is referred to as \msp. 
Unless otherwise noted, $t = 0$ denotes the start of the \pluto simulations.

\subsection{Phase II: Two-dimensional radiation-hydrodynamic simulations} \label{sec:met:pluto}
The surface density at 2.1\,Myr was used as the input for the second phase, where the model was evolved in axisymmetric 2D ($r$--$\theta$) spherical coordinates using \pluto to simulate the photoevaporative outflow and its impact on the disc. 
The photoevaporative wind driven by stellar high-energy irradiation was modelled following \citet{Picogna2019}, to which we refer for details.
We note that this model employs a line-of-sight heating approximation, which maximizes the effect of shadowing by neglecting diffuse heating; however previous work has demonstrated that the contribution of the diffuse radiation field is negligible in gap regions \citep{weber2024}.
The computational domain extended radially from $0.75$ to $380\,\mathrm{au}$, resolved with 512 logarithmically spaced radial cells.
The polar angle $\theta$ was discretized with 450 uniformly spaced cells between $0$ (the polar axis) and $\pi/2$ (the disc midplane).
Axisymmetry with respect to the azimuthal coordinate $\phi$ and mirror symmetry about the midplane were imposed. 
Although employing a symmetric midplane boundary can introduce numerical inaccuracies, this is not a concern in our simulations, as the flow near the midplane is dominated by Keplerian rotation with negligible vertical and radial velocity components, and the strong pressure gradients driving the wind are confined to the disc surface layers.
At the inner radial boundary, the azimuthal velocity was extrapolated assuming Keplerian rotation, with outflow conditions applied to the other variables.
To suppress numerical oscillations at the outer radial boundary, the damping zone described by \citep{Picogna2019} was adopted for $r > 360$\,au. 
At the polar axis, the dedicated \pluto polaraxis boundary condition was used, and equatorial symmetry was imposed at the midplane.
To prevent unphysical reflections at the outer radial boundary.

The surface density from the \msp model at 2.1\,Myr was mapped into the 2D domain assuming vertical hydrostatic equilibrium.
The initial radial and polar velocity components were set to zero, while the azimuthal velocity was initialized to Keplerian rotation.
This setup is referred to as model \mpg.

Since the \msp surface density profile at 2.1\,Myr already showed a pronounced depletion that could promote the formation of a deep gap in the 2D simulations, an additional \pluto run was performed using a modified initial surface density profile to suppress the early gap and assess its impact on gap opening.
For $R < 40\,\mathrm{au}$ the surface density was replaced by a power-law,
\begin{equation}
    f_\Sigma (R) = 10\,\mathrm{\frac{g}{cm^2}} \times (\frac{R}{10.4\,\mathrm{au}})^{-0.93},
\end{equation}
and the modified profile was defined as:
\begin{equation}
    \Sigma_{mod}(R) = max\bigl[\Sigma(R), f_\Sigma(R) \bigr].
\end{equation}
The powerlaw profile $f_\Sigma$ is shown as the solid red line in \autoref{fig:sigma0}.
This setup is referred to as model \mpng.

\section{Results}

\subsection{Gap evolution}

\begin{figure*}
	\centering
	\includegraphics[width=\textwidth]{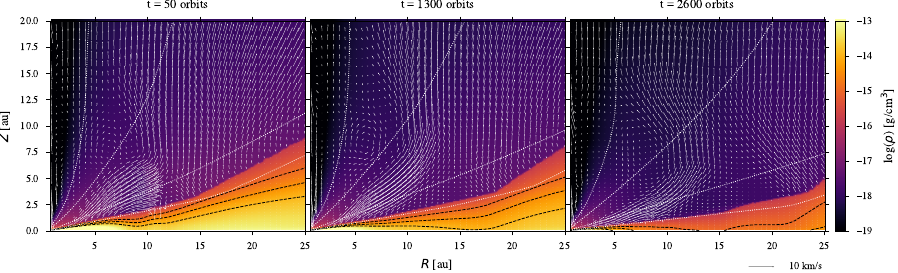}
	\caption[Global wind structure]{
		Gas density and velocity field of model \mpg after 50, 1300, and 2600 orbits. Each panel shows a time average over 10 orbits, taken over the intervals [40–50], [1290–1300], and [2590–2600] orbits, respectively. The white dotted lines represent column number density contours for values of (from top to bottom) $10^{19}, 10^{20}, 10^{21}$, and $10^{22}$\,cm$^{-2}$. The black dashed lines are density contours for values of (from top to bottom) $10^{-14}, 10^{-14.5}$, and $10^{-15}$\,g\,cm$^{-3}$.
	}
	\label{fig:wind-global}
\end{figure*}

\autoref{fig:wind-global} shows the evolution of the gas density and velocity field for model \mpg at three different epochs. No significant photoevaporative wind is launched at the location of the initial density depletion. Instead, the depleted region is partially replenished by a combination of viscous and pressure-driven radial flows, as well as material transported from neighbouring regions via the wind. At larger radii, the disc has a higher surface density, leading to a greater vertical extent of the optically thick disc. As a result, beyond a certain radius a substantial fraction of the outer disc extends above the shadow cast by the inner disc and, at the surface, becomes directly exposed to stellar irradiation. At this transition point, the wind is relaunched and the disc surface steepens markedly relative to the inner region.

As the system evolves, the transition point shifts outward with time. However, a low-density region persists between the inner disc and the outer wind zone, remaining in the shadow of the inner disc throughout the simulation. In this shadowed region, the local wind-driven mass loss is strongly reduced, leading to the formation of a partial gap rather than a fully evacuated one. This behaviour indicates that the photoevaporative flow dynamically responds to changes in the disc structure rather than maintaining the fixed radial dependence commonly assumed in one-dimensional models.
%* \autoref{fig:wind-global}: density and velocity field of model \mpg. No wind is launched at the location of the gap, instead the gap is partially replenished (viscously and through the wind). Moving to larger radii, the surface density increases again so that at some point enough material is in the upper layers of the disc, high enough not to be directly irradiated and not shadowed by the inner disc. At this "transition point" the wind starts to be launched again and the disc surface becomes much steeper than in the inner part of the disc. Over time, the transition point moves to larger radii, but between it and the inner disc remains always a lower-density region that lies in the shadow of the inner disc \\

%* \autoref{fig:sigma-evol}: comparison between runs with and without initial gap: the model without the initial gap converges to a similar gap profile (blue profile at 2600 orbits is similar to black profile at 800 orbits). differences in the inital mass mean slower evolution for the model without a gap. but this shows that even though we imposed a potentially unphysical, sharp gap, the long-term behaviour is independent of that. we can thus continue our analysis with the \mpg model, which is loosely representative of later evolutionary stages of the \mpng model \\

\begin{figure}
	\centering
	\includegraphics[width=.48\textwidth]{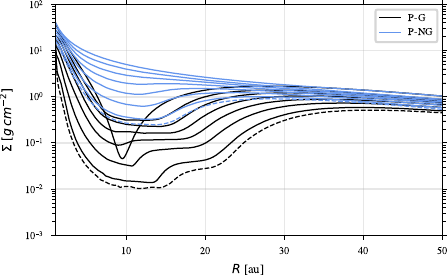}
	\caption[1D surface density evolution]{
		1D surface density evolution of model \mpg with an initial depletion (black lines) and \mpng without an initial depletion (blue lines). The solid lines represent the surface density in 400 orbit intervals (from 0 at the top to 2400 orbits at the bottom). The dashed line is the surface density at the end of the simulation at 2600 orbits.
	}
	\label{fig:sigma-evol}
\end{figure}

\autoref{fig:sigma-evol} compares the surface density evolution of discs with an initial depletion (\mpg, black curves) and without an initial depletion (\mpng, blue curves). The model without an initial depletion gradually develops a similar density structure, converging to the \mpg profile after $\approx2600$ orbits (comparable to the \mpg model after $\approx800$ orbits). The slower evolution of the initially smooth disc results from its larger initial mass, which delays viscous depletion and the onset of the wind–disc interaction. Nevertheless, the similarity of the final surface density profiles demonstrates that the long-term outcome is largely independent of the morphology of the initial density depletion. We therefore focus the subsequent analysis on the \mpg model, which can be regarded as representative of the later evolutionary stages of the \mpng model.

\subsection{Wind mass loss profiles}
\label{sub:mlossprof}
\begin{figure*}
	\centering
	\includegraphics[width=\textwidth]{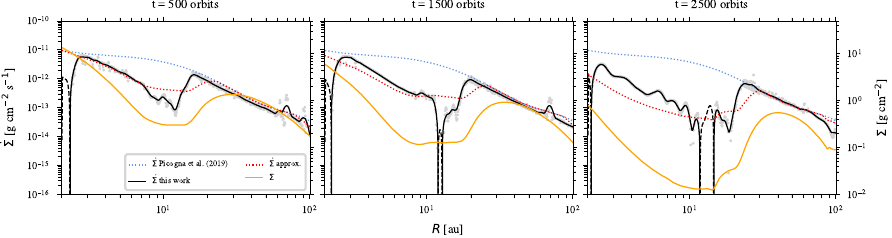}
	\caption[Wind mass loss profiles]{
		1D radial wind mass loss profiles after 500, 1500, and 2500 orbits. The blue dotted line is the profile for a primordial disc by \citet{Picogna2019}. The black solid line represents the mass loss profile measured in the model, smoothed with a Savitzky-Golay-Filter using second-order polynomials. The grey points show the exact measurements. The black dashed portions mean that the mass-loss is negative. The orange solid line represents the 1D surface density, and the red dotted line is our simple approximation as described in \autoref{sub:mlossprof}
	}
	\label{fig:sigmadot}
\end{figure*}

\autoref{fig:sigmadot} shows the radial profiles of the mass-loss rate, $\dot{\Sigma}$, at 500, 1500, and 2500 orbits, compared with the primordial disc profile of \citet{Picogna2019}. The mass-loss rate in our model is measured from the vertical flux through a surface located 0.3 au above the disc–wind interface, using density and velocity fields averaged over 10 orbits. At early times, the profile closely follows the primordial distribution, but as the partial gap develops, $\dot{\Sigma}$ exhibits a pronounced local minimum coinciding with the depleted region.

This depression in the mass-loss rate can be approximately reproduced by scaling the \citet{Picogna2019} profile according to the relative depth of the partial gap. The smooth, unperturbed surface density profile is estimated with a power law,

\begin{equation}
    \Sigma_\mathrm{smooth} = \Sigma_\mathrm{model}(R_\mathrm{max}) \bigl( \frac{R}{R_\mathrm{max}} \bigr)^{-1},
\end{equation}
where $\Sigma_\mathrm{model}$ is 1D the surface density in the model and $R_\mathrm{max}$ is the radius where $\Sigma_\mathrm{model}$ reaches its maximum.
This is then used to scale the original mass loss profile according to
\begin{equation}
    \dot{\Sigma}_\mathrm{approx} = max\bigl(1, \frac{\Sigma_\mathrm{model}}{\Sigma_\mathrm{smooth}}\bigr) \dot{\Sigma}_\mathrm{P19}.
\end{equation}

This prescription approximately captures the suppression of mass loss within the partial gap and its recovery at larger radii.

Inside $R \lesssim 2$ au, where the wind is launched nearly parallel to the disc surface, the measured values are unreliable and ignored. At large radii, the mass-loss rate converges toward the original \citet{Picogna2019} values, confirming that the wind remains unaffected far from the gap. The steep increase in $\Sigma$ at the outer edge of the gap correlates with a sharp rise in $\dot{\Sigma}$, though the measured enhancement exceeds that predicted by the approximation, indicating that the simple scaling underestimates the mass loss near the outer edge. Conversely, the inner edge is well reproduced. At later times, when the gap becomes deeper and wider ($t = 2500$ orbits), the approximation underestimates the total mass loss, likely because the inner disc no longer efficiently shields the outer regions from high-energy irradiation. In a narrow band within the most depleted parts of the partial gap, mass-loss rates are overestimated, and negative $\dot{\Sigma}$ values -- corresponding to inflow -- are occasionally present in the 2D models but cannot be captured by our approximation.

\begin{figure}
	\centering
	\includegraphics[width=.48\textwidth]{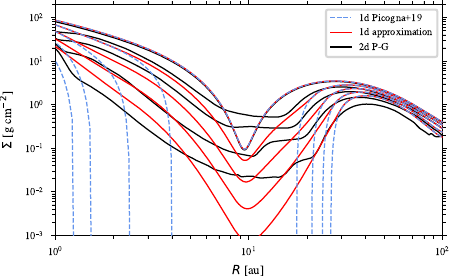}
	\caption[Modified surface density evolution in the \mpg model]{
		Surface density evolution of the 2D \mpg model without an initial depletion (\mpg; black), the corresponding 1D model with the modified mass-loss prescription (red), and the original 1D model with the mass loss profile by \citep{Picogna2019} (blue dashed). Lines of the same colour represent different timesteps at intervals of 25\,kyr starting at t=0 (top lines).
	}
	\label{fig:sigmacomp_pg}
\end{figure}

\autoref{fig:sigmacomp_pg} compares the surface density evolution of the original \mpg model (2D), the corresponding 1D model employing the modified mass-loss prescription, and the original \spock model employing the profile by \citet{Picogna2019}. While the inclusion of the modified mass-loss profile improves the agreement with the 2D results relative to the primordial prescription, the evolution remains noticeably faster and the resulting partial gap more pronounced than in the 2D simulation. This discrepancy likely arises from a combination of overestimated mass-loss rates within the depletion zone and high mass fluxes along and just above the disc surface in the 2D flow, which lie below the level at which the mass-loss rate is measured and act to partially refill and smooth the gap. Such radial mass transport cannot be captured by the simplified 1D approach, where only the vertical component of the wind is considered. Consequently, even the improved prescription cannot fully reproduce the self-regulated gap evolution seen in the multidimensional simulations.

\begin{figure}
	\centering
	\includegraphics[width=.48\textwidth]{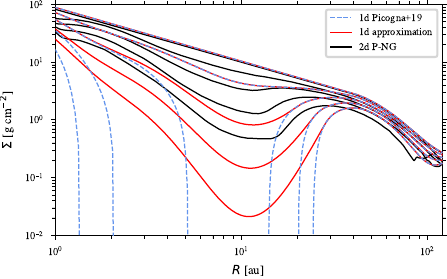}
	\caption[Modified surface density evolution in the \mpng model]{
		Surface density evolution of the 2D \mpng model without an initial depletion (\mpng; black), the corresponding 1D model with the modified mass-loss prescription (red), and the original 1D model with the mass loss profile by \citet{Picogna2019} (blue dashed).
        Lines of the same colour represent different timesteps at intervals of 25\,kyr starting at t=0 (top lines).
        The early agreement between the 1D and 2D profiles is somewhat improved relative to the \mpg case, but the 1D model still evolves faster and produces a more pronounced gap, underscoring the role of multidimensional mass transport in regulating disc structure. 
	}
	\label{fig:sigmacomp_png}
\end{figure}

A similar comparison for the \mpng model is shown in \autoref{fig:sigmacomp_png}. In this case, the agreement between the 1D and 2D evolutions is slightly improved at early times, since the simulation does not begin with an artificially sharp gap that would otherwise amplify the effects of unresolved   radial mass transport. Nonetheless, the 1D model again evolves too rapidly, producing a deeper and wider gap than the 2D simulation. As in the \mpg case, this discrepancy arises from the limitations of our approximation and the absence of high mass fluxes flowing along and just above the disc surface, which in the 2D flow act to redistribute material and moderate the contrast between the inner and outer disc.

%*\autoref{fig:sigmacomp_pg}: comparing the surface density evolution between the original spock model, the spock model with modified mass-loss profile, and the pluto model. it is clear that the modified profile is still not able to perfectly reproduce the 2d evolution (it evolves faster and still produces a more pronounced gap than expected). I believe this is mainly due to high mass-fluxes along and close to the disc surface that smear out the gap (TODO: try to visualize). these are not captured in the 1d profiles.\\
%* \autoref{fig:sigmacomp_png}: a bit better in the beginning, since we did not start with such a sharp gap that would be even more affected by what i explained above. nevertheless, it also evolves too quickly

%*TODO: include a figure comparing simple synthetic populations with both profiles? could also be published later with an improved approximation for the mass-loss profile that has been shown to work with different disc properties

\section{Discussion}
The results presented here show that for the adopted disc parameters, the ability of photoevaporation to open and maintain a gap remains uncertain. In our simulations, the mass loss in the gap region is strongly suppressed once the depletion begins, while  radial mass transport resulting from a combination of viscous torques, pressure gradients, and the wind acts to partially refill the gap. It is therefore unclear whether a fully evacuated cavity will eventually form, or whether the system will instead reach a quasi-steady state characterised by a persistent, but incomplete, partial gap. This implies that discs exhibiting moderate depletion without complete inner clearing may represent photoevaporatively influenced, but not fully dispersed, systems.

While our model explores a single set of disc and stellar parameters, the behaviour likely depends on the irradiation strength, disc viscosity, and mass. 

Even if the gas gap remains partially filled, the accompanying pressure maximum at its outer edge could still act as an efficient trap for dust grains \citep[e.g.][]{Pinilla_2012b}. Such localised dust accumulation may produce observational signatures resembling those of transition discs, even in the absence of complete inner-disc clearing. A detailed investigation of the dust evolution and the corresponding synthetic observables is beyond the scope of this work but will be essential to interpret high-resolution continuum observations of discs in the photoevaporative regime.

%Our results therefore challenge the prevailing paradigm in which photoevaporation is invoked as a robust mechanism for producing transition discs. Instead, the feedback between the evolving flow and the disc structure appears to limit the efficiency of gap opening, at least under the conditions explored here. Whether a gap can eventually form and clear entirely may depend sensitively on the initial disc mass, viscosity, irradiation level, and the strength of other wind-driving mechanisms.

Future work should also extend this analysis to a broader parameter space and explore the combined impact of photoevaporation and magnetically driven winds. Moreover, detailed radiation-transfer post-processing of these models will be necessary to establish the observable characteristics of partially depleted, shadowed discs and to assess whether such systems can indeed account for the diversity of observed transition-disc morphologies.

Finally, An important aspect of our method is the assumption that the gas is in radiative equilibrium, which enables us to decouple the radiative transfer calculation from the hydrodynamic evolution. Although this assumption may break down in the uppermost regions of the wind, it is expected to hold near the wind base, where the kinematical structure—both in velocity and density—is not markedly different from that obtained in primordial disc models. In particular, Owen et al. (2012) and Picogna et al. (2019) carried out a posteriori tests on their simulation grids and demonstrated that the gas in the relevant launching and flow-shaping regions is largely in radiative equilibrium. Given the similarity between their kinematics and those found in our simulations, their results provide supporting evidence that the radiative equilibrium approximation remains valid in the regime explored here.

\section{Conclusions}

We have investigated whether photoevaporation alone can open and sustain gaps in protoplanetary discs by following the coupled evolution of the flow and the disc structure in two-dimensional radiation–hydrodynamical simulations. Our results show that as a gap develops, the local mass-loss rate drops substantially, suppressing further clearing of the gap region. The resulting structure is continuously replenished by radial mass transport in the disc and through the wind. As a consequence, the system does not evolve toward a fully evacuated cavity, but rather maintains a persistent, partially depleted zone whose depth and width evolve only slowly over time.

A comparison between models with and without a pre-existing density depletion demonstrates that the long-term evolution is largely insensitive to initial morphology. In both cases, photoevaporation self-regulates the flow in such a way that the depletion saturates rather than proceeding to full clearing. This behaviour challenges the conventional picture in which photoevaporation is assumed to carve clean inner holes in the gas distribution and thus directly produce transition discs. Nevertheless, the associated pressure maximum at the outer edge of the depression may still provide an efficient trap for dust, leading to observational signatures consistent with transition-disc morphologies.

We also provide a first-order prescription to approximate this behaviour within one-dimensional disc evolution models by scaling the primordial mass-loss profile according to the local depletion of the gas surface density. While this approach improves upon existing static photoevaporation prescriptions, it remains an imprecise representation of the inherently multidimensional flow and cannot capture key effects such as radial mass redistribution or shadowing.

Further work is required to refine this prescription and to explore a wider parameter space in stellar mass, disc mass, viscosity, and irradiation properties. Incorporating such feedback-regulated photoevaporation into disc evolution frameworks will be key to assessing the interplay between internal winds and planet formation pathways. Extending this effort will be essential to establish reliable recipes for inclusion in planet formation and population synthesis models, ultimately linking disc dispersal physics with the emerging demographics of planetary systems.

\section*{Acknowledgements}

We thank the anonymous referee for a constructive report that helped to improve the clarity of our letter.
We acknowledge
support of the DFG (German Research Foundation), Research Unit “Transition discs” - 325594231 and of the Excellence Cluster ORIGINS - EXC-2094 - 390783311.
We acknowledge support from the Max Planck Society. Computations were performed on the HPC system CCAS at the Max Planck Computing and Data Facility. 

%%%%%%%%%%%%%%%%%%%%%%%%%%%%%%%%%%%%%%%%%%%%%%%%%%
\section*{Data Availability}

 The data files from the hydrodynamical simulations are available from the authors upon request.



\bibliographystyle{mnras}
\bibliography{references.bib} % if your bibtex file is called example.bib

%%%%%%%%%%%%%%%%%%%%%%%%%%%%%%%%%%%%%%%%%%%%%%%%%%

%%%%%%%%%%%%%%%%% APPENDICES %%%%%%%%%%%%%%%%%%%%%

%\appendix

%%%%%%%%%%%%%%%%%%%%%%%%%%%%%%%%%%%%%%%%%%%%%%%%%%

% Don't change these lines
\bsp	% typesetting comment
\label{lastpage}
\end{document}